\documentstyle[aps,multicol,epsfig]{revtex}
\newcommand{\nc}{\newcommand}
\nc{\be}{\begin{equation}}
\nc{\ee}{\end{equation}}
\nc{\bea}{\begin{eqnarray}}
\nc{\eea}{\end{eqnarray}}
\nc{\bean}{\begin{eqnarray*}}
\nc{\eean}{\end{eqnarray*}}
\nc{\mb}{\mbox}
\nc{\rnc}{\renewcommand}
\nc{\r}{\mb{\boldmath$r$}}
\nc{\x}{\mb{\boldmath$x$}}
\nc{\A}{\mb{\boldmath$A$}}
\nc{\sa}{\mb{\boldmath$a$}}
\nc{\sss}{\mb{\boldmath$\sigma$}}
\nc{\nab}{\nabla}
\nc{\X}{\sf x}

\renewcommand{\narrowtext}{\begin{multicols}{2} \global\columnwidth20.5pc}
\renewcommand{\widetext}{\end{multicols} \global\columnwidth42.5pc}

\begin{document}
\draft

\title{ Staggered flux state of electron in two-dimensional t-J model }
\author{Koichi Hamada and Daijiro Yoshioka}

\address{Department of Basic Science, 
University of Tokyo, 3-8-1 Komaba, Tokyo 153-8902}
\date{\today}
\maketitle

\begin{abstract} 

The competition between the staggered flux state, 
or the d-density wave state, 
and the d-wave pairing state is analyzed 
in two-dimensional $t$-$J$ model 
based on the U(1) slave boson mean-field theory.
Not only staggered flux of spinon but also staggered flux of holon 
are considered. 
In this formalism, 
the hopping order parameter of $physical$ electron is described 
by the product of hopping order parameters of spinon and holon.
The staggered flux amplitude of electron is the difference of 
staggered flux amplitude of spinon and that of holon.
In $\pi$-flux phase of spinon, staggered fluxes of spinon and holon  
cancel completely 
and staggered flux order of electron does not exist.
However, in staggered flux phase of spinon 
whose staggered flux amplitude is not $\pi$, fluxes 
does not cancel completely
and staggered flux amplitude of electron remains. 
Thus, the phase transition between these two phases, 
$\pi$-flux phase and staggered flux phase of spinon, 
becomes a second order transition in $physical$ electron picture. 
The order parameter which characterizes this transition 
is staggered flux order parameter of electron. 
A mean-field phase diagram is shown. 
It is proved analytically that 
there is no coexisistence of staggered flux and d-wave pairing.
The temperature dependences of Fermi surface 
and excitation gap at $(0,\pi)$ are shown.
These behaviors are consistent with angle-resolved photoemission 
spectroscopy (ARPES) experiments.
\end{abstract}


%\narrowtext

\begin{multicols}{2}
%\narrowtext

%%%%%%%%%%%%%%%%%%%%%%%%%%%%%%%%%%%%%%%%%%%%%%%%%%%%%
\section{Introduction}

Recently the staggered flux state\cite{AM88,HMA91} has been revived as 
a candidate of pseudo-gap phase of high $T_{\mathrm c}$ superconductors.    
The staggered flux state is characterized by the staggered orbital-current. 
Time-reversal-symmetry is broken 
if staggered current is the current of $physical$ electron. 
Weak magnetic signals were caught in recent experiments of 
underdoped YBa$_2$Cu$_3$O$_{6+x}$ (YBCO)  
by neutron scattering\cite{Mook01,Mook0204}. 
Static alternating magnetic signals were also caught by  
muon spin rotation experiment in the vortex cores of underdoped 
YBCO\cite{Miller}.
%These are observed neutron experiment 
The staggered flux state contains a density wave ordering 
whose symmetry is $d_{x^2-y^2}$\cite{Nayak00}.
Because of this, it is also called d-density wave. 
Thus, there is the gap whose symmetry is $d_{x^2-y^2}$.
Although $d_{x^2-y^2}$-gap exists in both 
the staggered flux state and the $d_{x^2-y^2}$-wave pairing state,   
the structures of the fermion excitation in these states are different.
The staggered flux state should have a segment-like Fermi-surface 
as observed in angle-resolved photoemission 
spectroscopy (ARPES) experiments
\cite{Loser96,Marshall96,Ding96,Harris96,Harris97,Ding97,Norman98,Yoshida0206}. 
On the contrary, $d_{x^2-y^2}$-wave pairing state has a point-like Fermi-surface.
The absence of specific heat anomaly may be thought 
to deny the staggered-flux phase.
However, we remark that there are arguments 
to justify the absence \cite{KeeKim02,Chakravarty0206}.
The lack of anomaly is either because of finite chemical potential\cite{KeeKim02},
or because that this transition can be described 
by 6-vertex model\cite{Chakravarty0206}.

The high-$T_{c}$ superconductors are doped Mott-insulators. 
Essential physics of these materials comes from strong Coulomb repulsion.
This physics is described by $t$-$J$ model\cite{Anderson87,ZhangRice88}.
A method that restrict double occupancy is the slave boson method. 
In both the slave boson methods, the U(1) method and the SU(2) method,  
the physical electron operator is described by a product of auxiliary 
fermions and bosons. 
In the U(1) slave boson theory, 
the physical electron operator $c_{i\sigma}$ is described by a product of    
an auxiliary spin-1/2 neutral fermion operator $f_{i\sigma}$ called spinon  
and a auxiliary spinless charged boson operator $b_{i}$ called holon;  
$c_{i \sigma}=b^{\dag}_{i}f_{i\sigma}$ with a constraint  
$b_{i}^{\dagger}b_{i}+ f_{i\sigma}^{\dagger}f_{i\sigma}=1$.
Here, the repeated spin index $\sigma$ is summed up over the two spin states.
On the contrary, in the SU(2) slave boson theory\cite{WenLee96,Lee98}, 
two auxiliary bosons are introduced.
The physical electron operator $c_{i\sigma}$ is described by a product of
an auxiliary isospin-1/2 neutral fermion operator $f_{i\sigma}$ 
and two auxiliary spinless charged boson operators, $b_{i1}$,$b_{i2}$; 
$c_{i\sigma}= h^{\dag}_{i}\psi_{i\sigma}/ \sqrt{2}=
 ( b^{\dag}_{i1}f_{i\sigma}+ 
 b^{\dag}_{i2}\epsilon_{\sigma \sigma^{'}} f^{\dag}_{i \sigma^{'}} )/ \sqrt{2}$
with three constraints, 
$\frac{1}{2} \psi^{\dag}_{i\sigma} \mbox{\boldmath $\tau$} \psi_{i\sigma} + 
h^{\dag}_{i} \mbox{\boldmath $\tau$} h_{i}=0$.
Here, $h_{i}^T=(b_{i1},b_{i2})$, 
$\psi_{i}^T=(f_{i\sigma},\epsilon_{\sigma \sigma^{'}} f^{\dag}_{i\sigma^{'}})$, 
and $\mbox{\boldmath $\tau$}
=(\tau_1, \tau_2, \tau_3)$ are Pauli matrices.

Wen and Lee proposed based on the SU(2) theory \cite{Lee98,Lee_Wen01,Lee0201} that 
the pseudo-gap state is a mixture of staggered flux state 
and $d_{x^2-y^2}$ superconducting state.  
These orders are dynamical rather than static 
and that the time-reversal symmetry is not broken.  
They discussed this in the region where bosons are condensed, 
$(z_{i1},z_{i2}) \neq {\bf 0}$, 
based on the $O(4)$ sigma-model description.
It is also proposed by Lee and Wen\cite{Lee_Wen01} 
that the static staggered-flux order can only exists in the vortex core.  
Short comings of this theory are that finite size Fermi-surface 
is not obtained naturally and that the condensation of bosons is assumed. 
The staggered current of physical electron and that of spinon 
are not equivalent in the staggered flux\cite{AM88,HMA91} phase
at finite temperature 
because the staggered flux phase of spinon exists 
above the temperature of Bose-condensation of holon\cite{WenLee96,Lee98,UL92}.
Above the temperature of Bose-condensation, 
holon operators cannot be treated as classical numbers (c-number).\cite{BC} 
The correspondence between the electron current and the spinon current 
is not obvious.

On the contrary, Chakravarty, Laughlin, Morr, and Nayak \cite{CL01} 
proposed that 
the staggered flux order exists as a static form in the pseudo-gap state.  
They proposed this at zero temperature based on Ginzburg-Landau theory.
The staggered flux state they proposed is that of electron and 
time-reversal symmetry is broken. 
They also proposed that there exist three ground states. 
The ground state changes as doping increases from 
pure staggered flux phase to 
the coexisting phase of staggered flux state and d-wave superconductivity(dSC) state 
and finally to pure dSC phase.
A short comings of their theory is that it is a phenomenology 
without microscopic foundation. 

In this paper, we analyze 
the staggered-flux state of physical electron in 
two-dimensional $t$-$J$ model by the U(1) slave boson method  
without assuming the Bose condensation of holons.
The relation between the staggered current of electron and 
the staggered current of  spinon is provided.
In Sec. II, our formulation will be reviewed. 
Not only an auxiliary field $\chi_{ij}$ that describes spinon-hopping 
but also  an auxiliary field $B_{ij}$ that describes holon-hopping  
will be introduced to decouple the hopping Hamiltonian. 
The advantages of this formulation will be stated 
in Sec. II A. 
For mean field solution, 
not only staggered flux of spinon but also staggered flux of holon 
will be considered. 
This solution provides the relation between the staggered flux state 
of spinon and that of electron.
In Sec. III, 
The self-consistent equations will be derived.   
In Sec. IV, the phase diagram is shown. 
where  the staggered-flux phase of electron exists. 
In Appendix,
it is proved analytically 
that the staggered flux and the d-wave pairing do not coexist.
The instability of these two states, 
the staggered flux state and the d-wave pairing state,  
are discussed. 
In Sec. V, the temperature dependences of 
Fermi surface and excitation gap at $(0,\pi)$ are shown.
We show that the Fermi-surface in the staggered-flux state resembles 
segment-like Fermi-surface observed by ARPES experiments. 
Finally, conclusion will be given in Sec. VI.  
Part of the present results will be published elsewhere\cite{LT23}.

\section{formalism}
\subsection{Slave boson t-J model and  auxiliary fields}

The starting Hamiltonian is 
two-dimensional t-J model on a square lattice\cite{ZhangRice88},  
\bea
H = \ -t \sum_{ \langle i,j \rangle }\mbox{\boldmath $P$}(c_{i\sigma}^{\dagger}c_{j\sigma}+h.c.) \mbox{\boldmath $P$}
+ J \sum_{\langle i,j \rangle} \mbox{\boldmath $S$}_{i}\cdot\mbox{\boldmath $S$}_{j}.
\eea
Here, $\langle i,j \rangle$ represents sum over the  nearest-neighbor sites, 
the repeated spin index $\sigma$ is summed up over the two spin states, and
{\mbox {\boldmath $P$} } is a projection operator to no doubly occupied state,  
$\mbox{\boldmath $S$}_i=\frac{1}{2}c_{i\sigma}^{\dagger}(\mbox{\boldmath $\sigma$})_{\sigma \sigma^{'}}c_{i\sigma^{'}}$,  
where \mbox{\boldmath $\sigma$}$=(\sigma_1, \sigma_2, \sigma_3)$ are Pauli matrices.
One way to represent the projection to no double occupancy is 
the slave boson method.
In the U(1) slave boson method, 
the electron operator $c_{i\sigma}$ is described by a product of    
an auxiliary spin-1/2 neutral fermion operator $f_{i\sigma}$ called spinon 
and an auxiliary spinless charged boson operator $b_{i}$ called holon; 
$c_{i \sigma}=b^{\dag}_{i}f_{i\sigma}$ with a constraint  
$b_{i}^{\dagger}b_{i}+ f_{i\sigma}^{\dagger}f_{i\sigma}=1$.
Here, the repeated spin index $\sigma$ is summed up over the two spin states.
In the path integral formalism, 
the partition function 
$Z(\beta)={\mathrm Tr} \exp (-\beta H )$   
is described by following functional integral, 
$Z=\int [db][df][d \lambda] \exp(-S)$, 
\bea
S&=&\int_0^\beta d\tau
\big[ \sum_i ( b^{\dag}_i \partial_\tau b_i 
+ f^{\dag}_{i\sigma}\partial_\tau f_{i\sigma} ) + H \big],  
\\
H &=& \ -t \sum_{ \langle i,j \rangle} 
(f_{i\sigma}^{\dagger}f_{j\sigma}b_{j}^{\dagger}b_{i}+c.c.)
+ J \sum_{ \langle i,j \rangle } \mbox{\boldmath $S$}_{i}\cdot\mbox{\boldmath $S$}_{j}
\nonumber \\
&&+ i \sum_{i} \lambda_{i} (b_{i}^{\dagger}b_{i}+ f_{i\sigma}^{\dagger}f_{i\sigma} - 1 )
-\mu_e \sum_{i} f^{\dag}_{i\sigma} f_{i\sigma}, 
\label{Z:1}
\eea
where $\beta$ is the inverse temperature $\beta=1/T$, 
$\tau$ is the imaginary time, 
$\int [db]= \int \prod_{i,\tau}db^{\dag}_i(\tau)db_{i}(\tau)$ 
is a complex boson integral, 
$\int [df]= \int \prod_{i,\tau, \sigma }df^{\dag}_{i \sigma}(\tau)df_{i  \sigma}(\tau)$ 
is a complex Grassmann integral, 
$\int [d\lambda]= \int \prod_{i,\tau}d\lambda_{i}(\tau)$ 
is the Lagrange-multiplier integral that represents a constraint 
$b_{i}^{\dagger}b_{i}+ f_{i\sigma}^{\dagger}f_{i\sigma}=1$, 
 $\mbox{\boldmath $S$}_i
=\frac{1}{2}f_{i\sigma}^{\dagger}(\mbox{\boldmath $\sigma$})_{\sigma \sigma^{'}}f_{i\sigma^{'}}$, 
$\mu_e$ is chemicsl potential.
%, and $\lambda_{i}(\tau)$ represents the constraint of no double occupancy.
We introduce three complex auxiliary fields $\chi_{ij}$, $\eta_{ij}$, 
and $B_{ij}$ on links to decouple the Hamiltonian.

\bea
H^{'}=\sum_{<ij>}\big[
%\tilde{H}=\sum_{<ij>}\big[
&&-t\big\{ B_{ij} f^{\dag}_{j\sigma} f_{i\sigma}
+ \chi^{*}_{ij} b^{\dag}_{i} b_{j} 
\big\} +c.c.
\nonumber\\
&&+ t \big\{ B_{ij} \chi^{*}_{ij} + B^{*}_{ij} \chi_{ij} \big\} 
\nonumber\\
&&-\frac{3J}{8}\big\{ 
\chi_{ij} f^{\dag}_{j\sigma} f_{i\sigma}
+\eta^{*}_{ij} f_{j\sigma} f_{i\sigma^{'}} \epsilon_{\sigma \sigma^{'}}
\big\} +c.c.
\nonumber\\
&&+\frac{3J}{8}\big\{ |\chi_{ij}|^{2} + |\eta_{ij}|^{2} \big\}  \big]
\nonumber\\
+ i \sum_{i}&& \lambda_{i} (b_{i}^{\dagger}b_{i}+ f_{i\sigma}^{\dagger}f_{i\sigma} - 1 )
-\mu_e \sum_{i} f^{\dag}_{i\sigma} f_{i\sigma}.
\label{MFA}
\eea
Then partition function $Z(\beta)$ is rewritten as;
$Z = \int [db][df][d\lambda][d\chi][d\eta][dB] \exp(-S^{'})$, where 
$ S^{'}=\int_0^\beta d\tau
\big[ \sum_i ( b^{\dag}_i \partial_\tau b_i 
+ f^{\dag}_{i\sigma}\partial_\tau f_{i\sigma} ) + H^{'} \big]$. 
The following relations 
are obtained by differentiating the integrand in Eq. (\ref{MFA}) by 
$\chi^{*}_{ij}, \eta^{*}_{ij}$, and $B^{*}_{ij}$\cite{comment_identity}, 
\bea
{\bar \chi_{ij}} 
&\equiv& \langle \chi_{ij} \rangle
= \langle f_{i\sigma}^{\dag}f_{j\sigma} \rangle, 
\nonumber \\
{\bar \eta_{ij}} 
&\equiv & \langle \eta_{ij} \rangle
=\langle f_{i\uparrow}f_{j\downarrow} 
- f_{i\downarrow}f_{j\uparrow}\rangle, 
\nonumber \\
{\bar {B}_{ij} } 
&\equiv & \langle B_{ij} \rangle 
= \langle b^{\dag}_{i} b_{j} \rangle. 
\eea
According to these relations, 
$\chi_{ij}$ describes the hopping of spinon, 
$\eta_{ij}$ describes the singlet pairing of spinon 
which is called resonating-valence bond (RVB),
and $B_{ij}$ describes the hopping of holon.
The quantity $B_{ij}=\langle b^{\dag}_{i} b_{j} \rangle$ can have a finite value 
even if there is no Bose-condensation of holon(see Sec. IV).  

Present treatment is different from previous one by Ubbens and Lee\cite{UL92}. 
In Ref. [\ref{UL92}], $B_{ij}$ was not introduced. 
Then $\langle \chi_{ij} \rangle$ can not be interpreted as spinon hopping  
$\langle f^{\dag}_{i\sigma}f_{j\sigma} \rangle$, 
but the correct relation becomes 
 $\langle \chi_{ij} \rangle 
= \langle f^{\dag}_{i\sigma}f_{j\sigma} 
+ \frac{8t}{3J} b^{\dag}_{i}b_{j} \rangle$ 
as can be known from Eq. (\ref{MFA}) 
without $B_{ij}$ term. 
In their method, the four-boson term 
$b^{\dag}_{i}b_{j}b^{\dag}_{j}b_{i}$ 
is created by the decoupling, 
but it is neglected based on an argument 
that the effect of this term is small at low doping. 
The situation is similar in the SU(2) method\cite{WenLee96,Lee98}, 
if $B_{ij}$ is not introduced. 
Our treatment has a merit that the four-boson term does not appear 
in the decoupling.

\subsection{Saddle point solution}

\begin{center}
\begin{figure}[h]
\epsfxsize 6cm \epsffile{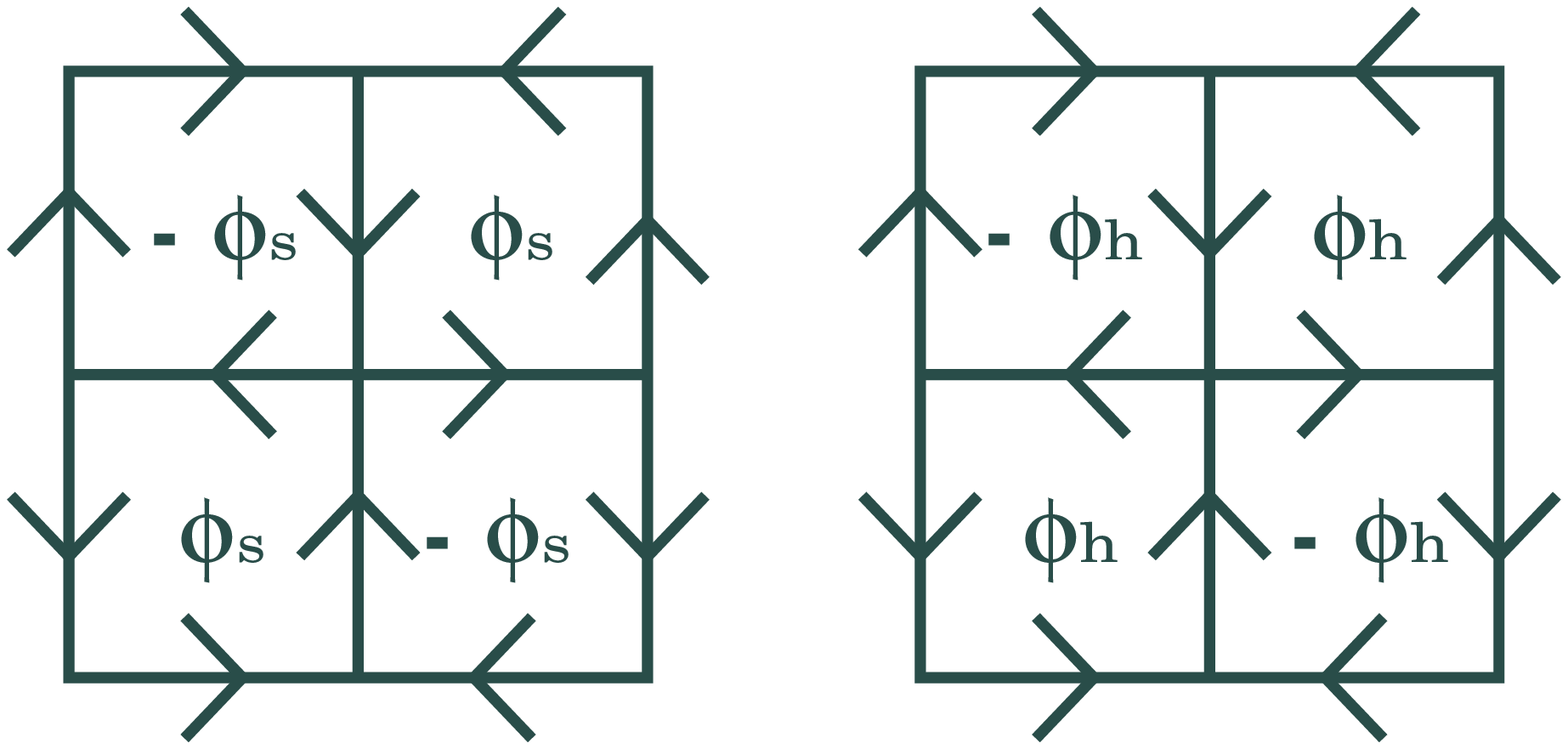}
\caption{ 
Staggered flux of spinon $\phi_{\mathrm s}$ and  holon $\phi_{\mathrm h}$.
}
\label{flux}
\end{figure}
\end{center}

We approximate the integral by auxiliary fields 
$\chi_{ij}$, $\eta_{ij}$, and $B_{ij}$ 
with their saddle point values 
$\bar{\chi}_{ij}$, $\bar{ \eta}_{ij}$, and $\bar{B}_{ij}$.
For the saddle point solution of $\chi_{ij}$ and $B_{ij}$,
we considered not only the staggered flux order 
of the spinon $\phi_{\mathrm s}$
but also the staggered flux order of the holon $\phi_{\mathrm s} $(Fig. \ref{flux}), 
namely

\bea
 {\bar \chi}_{i+ {\hat x},i}
&=& \chi {\mathrm e}^{i(-1)^{i} \phi_{\mathrm s}/4 }\ \ 
 = x_{\mathrm{s}} + i(-1)^{i} y_{\mathrm{s}}, 
\\
 {\bar \chi}_{i+ {\hat y},i}
&=&\chi {\mathrm e}^{-  i(-1)^{i} \phi_{\mathrm s}/4 } \ 
= x_{\mathrm{s}} - i(-1)^{i} y_{\mathrm{s}}, 
\\ 
 {\bar B}_{i+ {\hat x},i}
&=& B {\mathrm e}^{i(-1)^{i} \phi_{\mathrm h}/4 } \ 
= x_{\mathrm{h}} + i(-1)^{i} y_{\mathrm{h}}, 
\\
 {\bar B}_{i+ {\hat y},i}
&=& B {\mathrm e}^{-i(-1)^{i} \phi_{\mathrm h}/4 } 
= x_{\mathrm{h}} - i(-1)^{i} y_{\mathrm{h}}.
\eea

Here, ${\hat x}$ and ${\hat y}$ are unit vectors in the $x$ and $y$ direction,
$x_{\mathrm{s}}= \chi \cos(\phi_{\mathrm{s}}/4)$, 
$y_{\mathrm{s}}= \ \chi \sin(\phi_{\mathrm{s}}/4)$,  
$x_{\mathrm{h}}= B \cos(\phi_{\mathrm{h}}/4)$, 
$y_{\mathrm{h}}= B \sin(\phi_{\mathrm{h}}/4)$\cite{comment_choice}.
The staggered flux state contains a density wave (particle-hole pairing) ordering
whose symmetry is $d_{x^2-y^2}$, which is called ``d-density wave''
\cite{Nayak00}. 
The order parameters $y_{\mathrm{s}}$ and $y_{\mathrm{h}}$ correspond to the  
d-density wave 
%staggered flux 
order parameter of spinon and holon, 
respectively\cite{comment_ddw}.
For the symmetry of spinon pairing, 
we considered $d_{x^2-y^2}$\cite{Kotliar88,Suzumura88}, namely  
\bea
{\bar \eta}_{i+ {\hat x},i}= -\ {\bar \eta}_{i+ {\hat y},i}=\eta .
\eea

The integral by Lagrange multiplier field 
$\lambda_{i}$ is also approximated 
with its saddle point value $\bar{\lambda_{i}}=-i\lambda_{0}$. 
Then spinon and holon have chemical potentials, 
$\mu=\mu_{\mathrm e} - \lambda_0$ and
$\mu_{\mathrm h}= -\lambda_0$, respectively. 
These chemical potentials enforce the global constraints of 
spinon number and holon number, 
$\langle f^{\dag}_{i}f_{i} \rangle =1-\delta$ and 
$\langle b^{\dag}_{i}b_{i} \rangle =\delta$, respectively. 

This method that treats the spinon and the holon as separated degrees of freedoms  
is justified 
because the low temperature phase of the $t$-$J$ model is in the deconfinement phase 
of spinon and holon even when full gauge fluctuations
(fluctuations around this saddle point) are included. 
It is shown based on the compact U(1) treatments of the gauge fields
\cite{Nagaosa93,IM}. 
As the gauge field couples with finite density fermion,  
the gauge coupling gets weaker.
In the deconfinement phase of spinon and holon, 
the gauge field can be treated perturbatively.

Our formalism is an extension of the study by Ubbens and Lee \cite{UL92}.
The $f_{j\sigma}^{\dag}f_{i\sigma}$ term couples 
not only to $(3J/8)\chi_{ij}$ 
but also to $tB_{ij}$, 
namely 
the spinon feels the spinon staggered-flux($\phi_{\mathrm{s}}$) and
 the holon staggered-flux($\phi_{\mathrm{h}}$). 
The expectation values of the holon, $B$ and $\phi_{\mathrm{h}}$, 
have finite values  for the solution of self-consistency equations. 
Two advantages exist in our formalism;  
1) this is a new saddle point solution whose free energy is lower than the previous one, 
2) this solution provides the relation between the staggered-flux of electron and that of spinon.

With holon order parameters ${\bar B}_{ij}$, 
the hopping order parameter of  electron  becomes  
a product of the hopping order parameters of spinon and holon, 
\begin{eqnarray}
\langle c_{i\sigma}^{\dag}c_{j\sigma} \rangle
&=&\langle  f_{i\sigma}^{\dag} f_{j\sigma} \rangle 
\langle b_{j}^{\dag} b_{i} \rangle
=\ {\bar \chi}_{ij} {\bar B}_{ij}^{*}
\nonumber \\
&=&\chi B {\mathrm e}^{\pm i 
( \phi_{\mathrm s} - \phi_{\mathrm h}) /4 }.  
\end{eqnarray}
The electron staggered-flux order parameter $\phi_{\mathrm{e}}$ and 
the electron d-density wave order parameter $y_{\mathrm{e}}$ are given by 
\begin{eqnarray}
\phi_{\mathrm{e}}= \phi_{\mathrm{s}} -\phi_{\mathrm{h}}
\mbox{\ , \ }
y_{\mathrm{e}}=\chi B \sin ( \phi_{\mathrm{e}}/4 ) .
\label{eq:elddw}
\end{eqnarray}
According to eq.(\ref{eq:elddw}), the d-density wave order parameter of 
electron $y_{e}$ can also be written as,
$y_{\mathrm{e}}=
-i\frac{1}{2N}\sum_{k} (\cos{k_x} - \cos{k_y})
\langle c^{\dag}_{\bf k \sigma} c_{\bf k+Q \sigma} \rangle 
$.

\section{Self consistent equations}

This saddle point is described by the following 
partition function 
%mean-field Hamiltonian,  
$Z^{MF} = \int [db][df] \exp(-S^{MF})$, where 
$ S^{MF}=\int_0^\beta d\tau
\big[ \sum_i ( b^{\dag}_i \partial_\tau b_i 
+ f^{\dag}_{i\sigma}\partial_\tau f_{i\sigma} ) + H^{MF} \big]$, 
%\nonumber \\
\bea
H^{MF}&=&  H^{MF}_{\mathrm matter} -N ( \mu - \mu_h )
\nonumber \\
&& + 2N \big[ \frac{3J}{8}(x_{\mathrm s}^2 + y_{\mathrm s}^2 + \eta^2) 
+ 2t(x_{\mathrm s} x_{\mathrm h} + y_{\mathrm s} y_{\mathrm h}) \big], 
\label{eq:MF}
\end{eqnarray}
the spinon chemical potential is $\mu=\mu_{\mathrm e} - \lambda_0$, and 
the holon chemical potential is $\mu_{\mathrm h}= -\lambda_0$. 
The saddle point value of $i\lambda_i$ is $i\bar{\lambda}= \lambda_0$.
Here, the matter part of this Hamiltonian is given by 
\begin{eqnarray}
H^{MF}_{\mathrm matter}&=& \sum_{{\bf k}} {}^{\prime}
 \Psi_{\bf k}^{\dagger} H_{\bf k}\Psi_{\bf k}
+\Psi_{\bf k}^{{\mathrm h} \dagger} H^{\mathrm h}_{\bf k}\Psi^{\mathrm h}_{\bf k}. 
\end{eqnarray}
where, $\sum_{{\bf k}}^{'}$  stands for a sum over the half Brillouin zone.
$\Psi_{\bf k}$ and $\Psi_{\bf k}^{\mathrm h}$ 
are vector representations of spinon and holon, 
$\Psi_{\bf k}^{\dagger}
= (f_{{\bf k}\uparrow}^{\dagger}, f_{{\bf k+Q}\uparrow}^{\dagger}, 
f_{{\bf -k}\downarrow}, f_{{\bf -k-Q}\downarrow})$,
$\Psi_{\bf k}^{{\mathrm h}\dagger}= 
(b_{{\bf k}}^{\dagger}, b_{{\bf k+Q}}^{\dagger} )$.
The matrices $H_{\bf k}$ and $H^{h}_{\bf k}$ are,
\begin{eqnarray}
 H_{\bf k}=\left( \begin{array}{cccc} 
 (\epsilon_{\bf k}-\mu) & iW_{\bf k} & \Delta_{{\bf k}} & 0 \\
  -iW_{{\bf k}} & -(\epsilon_{\bf k}+\mu) & 0 & -\Delta_{\bf k} \\
  \Delta_{\bf k} & 0 & -(\epsilon_{\bf k}-\mu) & iW_{{\bf k}} \\
   0 &-\Delta_{\bf k}  & -iW_{\bf k} & (\epsilon_{\bf k}+\mu) \\
    \end{array} \right), 
\end{eqnarray}
\begin{eqnarray}
   H_{\bf k}^{\mathrm h}=
\left( \begin{array}{cc} 
(\epsilon^{\mathrm h}_{\bf k}-\mu^{\mathrm h}) & iW^{\mathrm h}_{\bf k} \\
  -iW^{\mathrm h}_{{\bf k}} & -(\epsilon^{\mathrm h}_{\bf k}+\mu^{\mathrm h}) \\
    \end{array} \right),  
\end{eqnarray}
where
\bea
\epsilon_{{\bf k}} 
&=&- ( 2t x_{\mathrm h} + \frac{3J}{4} x_{\mathrm s} ) (\cos k_{x}+ \cos k_{y}),
\\
 W_{{\bf k}} 
&=& \ ( 2t y_{\mathrm h} + \frac{3J}{4} y_{\mathrm s} ) (\cos k_{x}- \cos k_{y}), 
\\
 \Delta_{{\bf k}} 
&=&  \ \frac{3J}{4} \eta \ (\cos k_{x}- \cos k_{y}),
\eea
\bea
\epsilon^{\mathrm h}_{{\bf k}} 
&=&-2 t x_{\mathrm s}(\cos k_{x}+ \cos k_{y}),
\\
W^{\mathrm h}_{{\bf k}}
& =& 2 t y_{\mathrm s}(\cos k_{x}- \cos k_{y}).
\eea

The spinon couples not only to spinon order parameters, $x_{\mathrm s}$ and $y_{\mathrm s}$,   
but also to holon order parameters, $x_{\mathrm h}$ and $y_{\mathrm h}$.
The holon couples only to spinon order parameters, $x_{\mathrm s}$ and $y_{\mathrm s}$.
There exist two components, $W_{\bf k}$ and $\Delta_{\bf k}$, 
that generate the $d_{x^2-y^2}$ wave gap.  

After the Bogoliubov transformation, diagonalized Hamiltonian is obtained,
\bea
H^{MF}_{\mathrm matter}
&= &\sum_{{\bf k},s
=\pm 1} {}^{\prime}
E_{ {\bf k} s} 
( \alpha^{\dag}_{ {\bf k} s } \alpha_{ {\bf k} s } 
- \beta^{\dag}_{{\bf k}s} \beta_{ {\bf k} s } )
+E^{\mathrm h}_{ {\bf k} s}  
\alpha^{{\mathrm h} \dag}_{{\bf k}s} \alpha^{\mathrm h}_{ {\bf k} s }
\eea
where 
\bea
E_{ {\bf k} s}
&=& \sqrt{ ( s \sqrt{\epsilon_{{\bf k}}^{2}+W_{{\bf k}}^{2}}-\mu)^{2}+\Delta_{{\bf k}}^{2}},
\\
E^{\mathrm h}_{{\bf k} s}
&=& s \sqrt{ \epsilon_{{\bf k}}^{{\mathrm h}2} 
+ W_{{\bf k}}^{{\mathrm h}2} } -\mu^{\mathrm h} ,
\eea
Both $\alpha_{ {\bf k} s }$ and $\beta_{ {\bf k} s }$ are fermionic fields, 
$\alpha^{\mathrm h}_{ {\bf k} s }$ is bosonic field, 
and $E^{( {\mathrm h} )}_{ {\bf k} s}$ is excitation spectrum of spinon (holon).
Index $s$ describes the band index, which takes the value $+1$ or $-1$.
After integrating out  spinon field and  holon field, 
the free energy has the following form,
\bea
F=&& -2T \sum_{{\bf k}, s}{}^{\prime} \ln \cosh ( \beta E_{{\bf k} s} /2)
\nonumber \\
&& +T \sum_{{\bf k}, s}{}^{\prime} 
\ln (1- {\rm e}^{-\beta E^{\mathrm h}_{{\bf k} s} })
\nonumber \\
&& + 2N \big[ \frac{3J}{8}(x_{\mathrm s}^2 + y_{\mathrm s}^2 + \eta^2) 
+ 2t(x_{\mathrm s} x_{\mathrm h} + y_{\mathrm s} y_{\mathrm h}) \big]
\nonumber \\
&& -N {\mathrm \delta} ( \mu - \mu^{\mathrm h} ).
\eea
By minimizing the free energy $F$, we obtain the self-consistency equations,
\bea
x_{\mathrm s}&=& ( 2tx_{\mathrm h} + \frac{3J}{4} x_{\mathrm s} ) 
\frac{1}{N} \sum_{{\bf k}, s} {}^{\prime} 
\gamma_{{\bf k}+}^{2} 
\nonumber \\
&& \times \frac{ \tanh (\beta E_{{\bf k}s}/2) }{2 E_{{\bf k},s}} 
\Big[ 1 + \frac{s(-\mu)}{( \epsilon_{ {\bf k} }^2 + W_{ {\bf k} }^2 )^{1/2} } \Big], 
\\
y_{\mathrm s}&=& ( 2ty_{\mathrm h} + \frac{3J}{4} y_{\mathrm s} ) 
\frac{1}{N} \sum_{{\bf k}, s}{}^{\prime} 
\gamma_{{\bf k}-}^{2} 
\nonumber \\
&& \times \frac{ \tanh (\beta E_{{\bf k}s}/2) }{2 E_{{\bf k} s}} 
\Big[ 1 + \frac{s(-\mu)}{( \epsilon_{ {\bf k} }^2 + W_{ {\bf k} }^2 )^{1/2} } \Big], 
\label{ys}
\\
\eta&=& \frac{3J}{4} \eta \  \frac{1}{N} 
\sum_{{\bf k}, s}{}^{\prime} \gamma_{{\bf k}-}^{2} 
\frac{\tanh (\beta E_{{\bf k}s}/2) }{2 E_{{\bf k}s}}, 
\label{eta}
\\
x_{\mathrm h}&=& 2t x_{\mathrm s}  \frac{1}{N}
\sum_{{\bf k},s}{}^{\prime} \frac{\gamma_{{\bf k}+}^2}
{ 2 ( \epsilon_{ {\bf k} }^{{\mathrm h}2} + W_{ {\bf k} }^{{\mathrm h}2})^{1/2}}
\frac{(-s)}{ {\rm e}^{\beta E^{\mathrm h}_{{\bf k} s}} -1}, 
\\
y_{\mathrm h}&=& 2t y_{\mathrm s}  \frac{1}{N}
\sum_{{\bf k}, s}{}^{\prime} 
\frac{\gamma_{{\bf k}-}^2}{ 2 ( \epsilon_{ {\bf k} }^{{\mathrm h}2} 
+ W_{ {\bf k} }^{{\mathrm h}2})^{1/2}}
\frac{(-s)}{ {\rm e}^{\beta E^{\mathrm h}_{{\bf k} s}} -1}.
\label{yh}
\eea
The chemical potential $\mu$ and $\mu_{\mathrm h}$ are determined by
\bea
{\mathrm \delta } &=& \frac{1}{N}
\sum_{{\bf k}, s}{}^{\prime} 
\Big( s ( \epsilon_{ {\bf k} }^2 + W_{ {\bf k} }^2 )^{1/2} - \mu \Big)
\frac{ \tanh ( \beta E_{{\bf k}s}/ 2 ) }{ E_{{\bf k}s} }, 
\\
{\mathrm \delta} &=& \frac{1}{N}
\sum_{{\bf k}, s}{}^{\prime} 
\frac{ 1}{ {\rm e}^{\beta E^{\mathrm h}_{{\bf k} s}} -1}, 
\eea
where $\gamma_{{\bf k}\pm} = \cos k_x \pm \cos k_y$.

At half-filling, where spinon chemical potential $\mu$ is zero 
and holon is absent (${\bar B}_{ij}=0$, i.e. $x_h=y_h=0$), 
$E_{ {\bf k} s} = 
\sqrt{  \epsilon_{{\bf k}}^{2}+
(3J/4)^2 (y_{\mathrm s}^{2} + \eta^{2}) ( \cos k_x -\cos k_y)^2 }$, 
and both $y_{\mathrm s}$ and $\eta$ have the same self-consistent equations.
Thus, staggered flux state and d-wave pairing state are degenerate.
All states that have the same value of $y_{\mathrm s}^2+\eta^2$ 
are degenerate.
However, which state has the lower energy at finite doping is known 
only after actual minimization of the free energy is done.

%%%%%%%%%%%%%%%%%%%
\section{Phase diagram and staggered current}

\subsection{ Phase diagram }

\begin{center}
\begin{figure}
\epsfxsize 7cm 
\epsffile{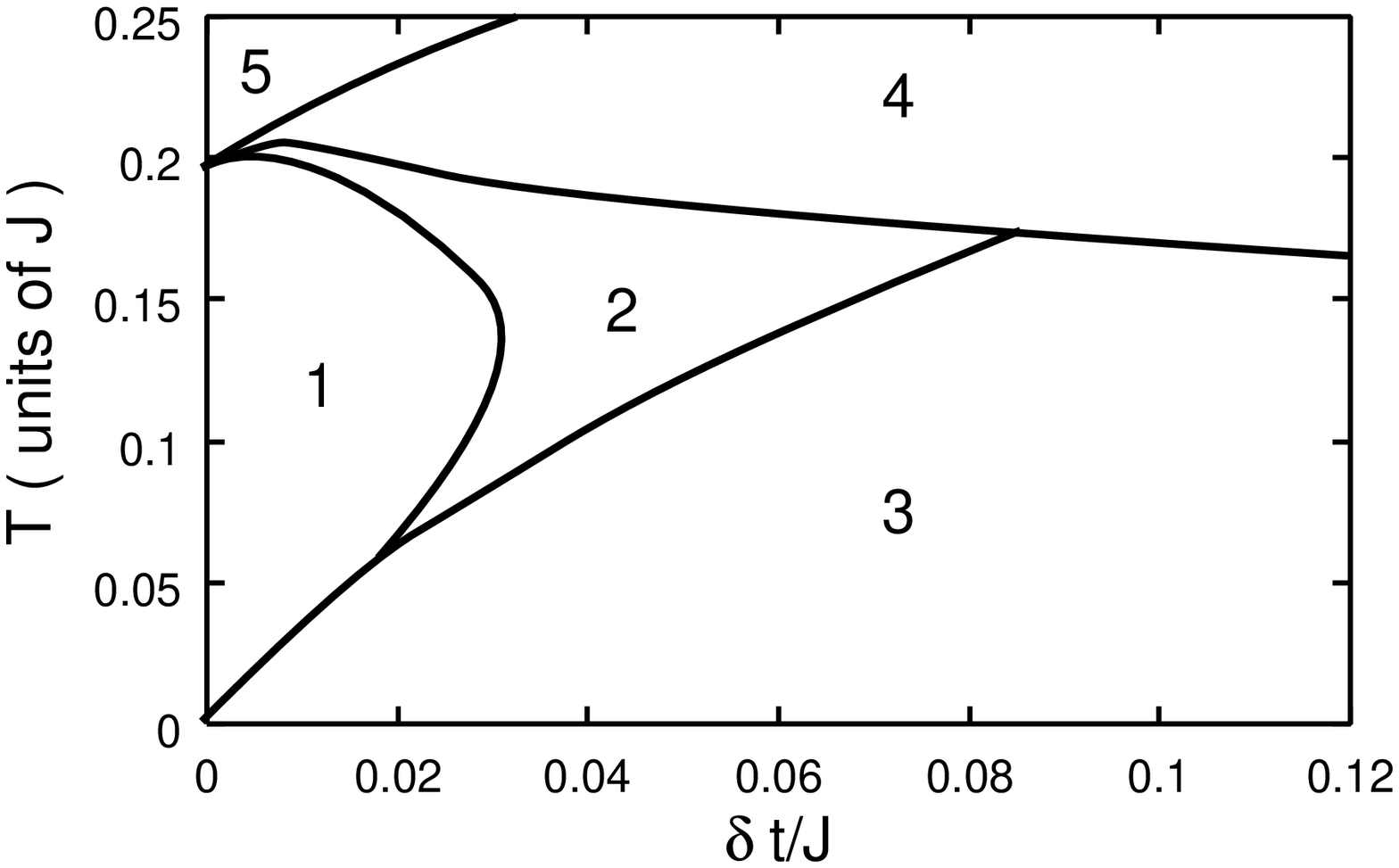}
\caption{  
Mean field phase diagram for $t/J=1$, 
where $\delta$ is hole concentration.
The staggered-flux order of electron exists only in region 2. 
In this phase, staggered-fluxes of spinon and holon 
do not cancel completely.
Thus the staggered flux of electron remains. 
In region 1, both spinon and holon are in the $\pi$-flux state. 
The fluxes cancel completely. 
The region 3 is the d-wave RVB phase where the $d_{x^2-y^2}$ pairing 
of spinon exists. 
The region 4 is the uniform RVB phase where hopping order parameters exist 
however it is real.
In the region 5, all order parameters are zero. 
The phase diagram for $t/J=2$ is quantitatively similar to the phase diagram 
for $t/J =1$.
}
\label{dgm}
\end{figure}
\end{center} 

 We solved the self-consistency equations numerically,
and obtained the phase diagram(Fig.\ \ref{dgm}) 
where
1) a region of the staggered-flux phase of the electron exists,
2) staggered-flux and d-wave pairing do not coexist, and
3) the ground state is a purely superconducting state.

 At half-filling, the holon order parameter ${\bar B}_{ij}$ is zero 
and the degeneracy of spinon states 
between the staggered-flux state and the d-wave pairing state exists 
due to local SU(2) symmetry\cite{Affleck88};
$\chi \ne 0, {\bar B}_{ij} =  0, y_{\mathrm{s}}^2 + \eta^2 = {\rm const} \ne 0$.
 In region 1, the staggered-flux of electron does not exist 
although the staggered flux of spinon exists. 
The spinon and holon states are both $\pi$-flux state\cite{comment_pi} respectively. 
In the electron picture, the staggered-flux is canceled completely;  
$\chi \ne 0, B \ne 0, \phi_{\mathrm{s}} = \phi_{\mathrm{h}}= \pi, \eta = 0$.
The d-density wave order parameter of the electron is 
$y_{\mathrm{e}}=\chi B \sin\big( (\pi - \pi)/4 \big) =0$.
The staggered-flux order of the electron exists only in region 2.
The staggered current of electron exists in this phase.
%can be observed experimentally.
The spinon staggered-flux $\phi_{\mathrm{s}}$ and holon staggered-flux 
$\phi_{\mathrm{h}}$ are not equal to $\pi$ nor 0, 
and the holon staggered-flux amplitude $\phi_h$ is not equal to 
the spinon staggered-flux amplitude $\phi_s$, 
$\phi_{\mathrm{s}} \ne \phi_{\mathrm{h}}$
; $\chi \ne 0, B \ne 0, \phi_{\mathrm{s}} \ne0, \phi_{\mathrm{h}} \ne 0, \eta = 0$, 
and $y_{\mathrm{e}}=\chi B \sin\big( (\phi_{\mathrm{s}} - \phi_{\mathrm{h}})/4 \big) \ne 0$.
 In region 3, $d_{x^2-y^2}$-wave pairing exists; 
$\chi \ne 0$ and $B \ne 0, \phi_{\mathrm{s}} = \phi_{\mathrm{h}} = 0, \eta \ne 0$.
and $y_{\mathrm{e}} = 0$.
 In region 4, there exists only uniform hopping order; 
$\chi \ne 0, B \ne 0, \phi_{\mathrm{s}} = \phi_{\mathrm{h}} = \eta = 0$, 
and $y_{\mathrm{e}} = 0$.
 In region 5, all order parameters are zero.
Spinon and holon cannot hop; 
$\chi = B = \phi_{\mathrm{s}} = \phi_{\mathrm{h}} = \eta = 0$.
and $y_{\mathrm{e}} = 0$.
The phase transitions, region 1 to region 3 and region 2 to region 3, 
are first order.
Other phase transitions are second order.
It is proved analytically 
that the staggered flux and the d-wave pairing do not coexist at finite doping.
Details are given in Appendix.

With the boson order parameter $B_{ij}$, 
the $\pi$-flux phase and staggered-flux phase of spinon and holon 
extends to higher-doped region compared to 
the previous work \cite{UL92}, where  $B_{ij}$ was not considered.
 The transition between region 1 and region 2   
is a second order transition in our theory.
(If one only focuses on spinon degree of freedom, 
this does not look like phase transition \cite{WenLee96,Lee98,UL92}.)
There exists an order parameter that characterizes this transition.
It is the staggered-flux of the electron.
This is shown in Sec. IV B. 

Although the ground state is the d-wave superconducting phase  
at finite doping, 
the staggered-flux phase exists at finite temperature.
The staggered-flux state has larger entropy 
than the d-wave pairing state, 
due to the larger number of the excitation 
around finite size Fermi-surface.
On the contrary, the Fermi-surface in the d-wave pairing state 
is always point-like. 
Therefore the staggered-flux state arises when temperature increases.
These are shown in Sec. V. 

When the fluctuation around the saddle point solution (gauge field) 
is included, 
it is expected that the staggered flux phase extends to still higher doped region
compared with the mean field phase diagram.
There are three reasons, 
1) the d-wave pairing of spinon will be destroyed, 
2) the instability to the staggered-flux state exists, 
and 3) the symmetry breaking of the staggered-flux order is discrete. 
Let us elaborate these reasons. 
Firstly, the d-wave pairing of spinon is destroyed 
above the Bose-condensation temperature of holon 
when the fluctuation around the saddle point solution (gauge field) 
is included\cite{UL94}. 
The d-wave RVB state without Bose-condenstation does not occur, 
and the transition to the d-wave pairing state becomes  
the direct transition to the d-wave superconducting state.  
The second reason is that the instability to the staggered-flux state 
has been discovered also in the gauge theory that includes 
the full lattice structure\cite{HPL,Kuboki94}.
It is to the staggered-flux state at low doped region 
and to the flux density-wave state (incommensurate staggered-flux state) 
at high doped region.
The momentum that has anomaly is ${\bf Q}=(\pi, \pi)$
to the staggered-flux state, 
and is $(\pi, \pi-\epsilon)$ 
to the flux density-wave.  
At finite temperature, these anomalies occur. 
The third reason 
is that staggered flux state is more robust against the fluctuation 
than the d-wave pairing (d-wave RVB) state.
This comes from the difference in the symmetry breakings. 
The symmetry breaking in the staggered-flux state is 
discrete symmetry, $Z_{2}$. 
On the other hand, continuous symmetry, $U(1)$,   
is broken in the d-wave pairing state. 

\subsection{Staggered-flux of the electron} 

\begin{center}
\begin{figure}[h]
\epsfxsize 7.2cm \epsffile{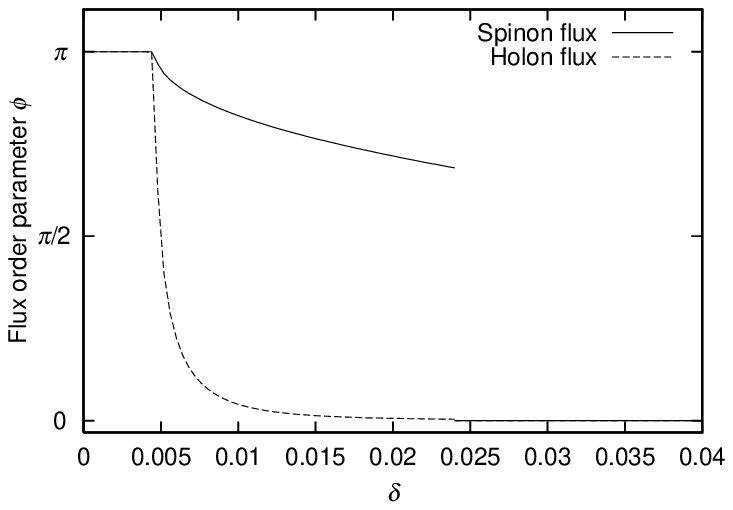}
\epsfxsize 7.2cm \epsffile{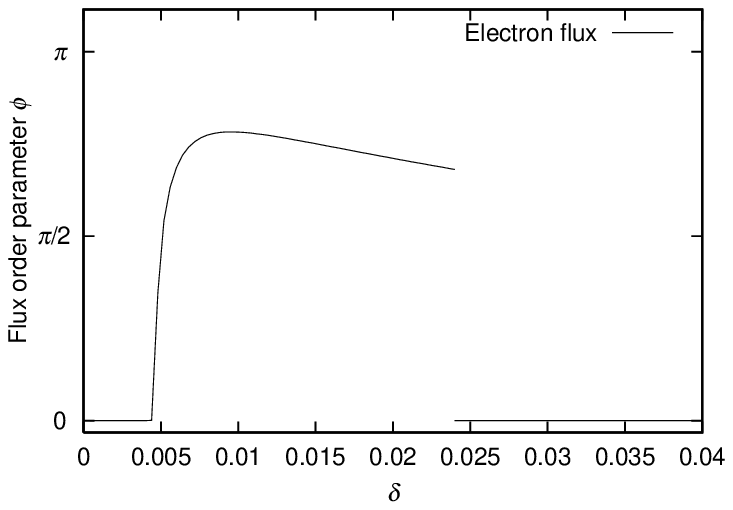}
\caption{ The staggered flux order parameter of spinon $\phi_{\mathrm s}$, 
holon $\phi_{\mathrm h}$, and electron $\phi_{\mathrm e}=\phi_{\mathrm s}-\phi_{\mathrm h}$ 
at $T=0.15J$ for $t/J=2$. $\delta$ is hole concentration.}
\label{s-flux}
\end{figure}
\end{center}

The doping dependences of the staggered-flux amplitudes  
are shown in Fig. \ref{s-flux}. 
The electron staggered-flux amplitude $\phi_{\mathrm e}$ is 
a difference of the staggered-flux amplitude 
of spinon $\phi_{\mathrm s}$ and holon $\phi_{\mathrm h}$ Eq.(\ref{eq:elddw}).
In Fig. \ref{s-flux}  the phase changes as doping increases 
from region 1 to region 2 and to finally region 3.
In region 1 which is $\pi$-flux phase of spinon, 
staggered-fluxes of spinon and holon cancel completely 
and staggered-flux order of electron does not exist.
However, in region 2, 
staggered flux amplitude is not $\pi$, 
fluxes does not cancel completely
and staggered-flux amplitude of electron remains. 
The staggered-flux amplitude of holon decays rapidly as doping increases 
but it is not zero.
At the transition point to d-wave pairing, staggred-flux order parameter becomes zero 
discontinuously. 

\subsection{Spontaneous current of the electron}

Spinon current and holon current cancel locally, 
\begin{eqnarray}
J^{\mathrm s}_{ij}  +  J^{\mathrm h}_{ij} = 0.  
\end{eqnarray}
The spinon current $J^{\mathrm s}_{ij}$ and  
the holon current $J^{\mathrm h}_{ij}$ are given by 
\begin{eqnarray}
J^{\mathrm s}_{ij} &=& 
 i \langle 
(\frac{3J}{8} {\bar \chi}^{*}_{ij}  + t {\bar B}^{*}_{ij} ) 
f^{\dag}_{i\sigma}f_{j\sigma}
- h.c. \rangle, 
\\
J^{\mathrm h}_{ij} &=& i 
t \langle
{\bar \chi}^{*}_{ij}  b^{\dag}_{i}b_{j} 
   - h.c. \rangle.
\end{eqnarray}
The $ \langle \bar{\chi}^{*}_{ij} f^{\dag}_{i} f_{j} \rangle $ term 
does not contribute to the current 
because it is real.
Only  $ \langle {\bar B}^{*}_{ij}  f^{\dag}_{i}f_{j} \rangle $ term 
contribute to the current.

The local cancellation of spinon current and holon current does not 
exclude the possibility of the electron staggered current.
In our formalism, 
the physical electron current is given by the imaginary part of 
the product of hopping order parameters of spinon ${\bar \chi}_{ij}$
and holon ${\bar B}^{*}_{ij}$.
\begin{eqnarray}
J^{\mathrm e}_{ij} 
&=& i t
\langle  c^{\dag}_{i} c_{j} -  h.c. \rangle
\nonumber \\
&=& it \big\{ 
\langle f^{\dag}_{i \sigma} f_{j \sigma} \rangle 
\langle  b^{\dag}_{j} b_{i} \rangle
- c.c. \big\}
\nonumber \\
&=& 2t \ {\mathrm Im} \big\{ 
{\bar \chi}_{ij} {\bar B}^{*}_{ij} 
\big\}
\nonumber \\
&=& 2t \ {\mathrm Im}
 \big\{ \chi B {\mathrm e}^{\pm i 
( \phi_{\mathrm s} - \phi_{\mathrm h}) /4 }  \big\}.
\end{eqnarray}
The explicit form of the electron current in this saddle point is 
$J^{\mathrm e}_{i+\hat{x} i} 
=(-1)^i2t \chi B \sin (  \phi_{\mathrm e} /4 ) 
= (-1)^i2t y_{\mathrm e}
$, $
J^{\mathrm e}_{i+\hat{y} i} 
=  -(-1)^i2t \chi B \sin (  \phi_{\mathrm e} /4 ) 
=  -(-1)^i2t y_{\mathrm e}
$.
The current of the electron is small at lower doped region because 
it is in proportion to the $B= | \langle b^{\dag}_{i} b_{j} \rangle |$. 
At low doping, $B$ is proportional to $\delta$ and small.  
The magnetic field due to the orbital current of electron is same order 
with previous estimations of $10^{-3}$T\cite{HMA91,CL01}.

%%%%%%%%%%%%%%%%%%%%%%%%%%%%%%%%%%%%
\section{Fermi surface and excitation gap at (0,$\pi$)}

The temperature dependences of Fermi surface 
and excitation gap at $(0,\pi)$ are shown.
These behaviors are consistent with angle-resolved photoemission 
spectroscopy (ARPES) experiments
\cite{Loser96,Marshall96,Ding96,Harris96,Harris97,Ding97,Norman98,Yoshida0206}.

\subsection{Fermi surface}

Fermi-surface is the loci of gapless excitation in momentum space.
It is defined by the zero-energy line of the electron excitation energy. 
The Fermi surface for $\delta=0.02$ is shown in Fig. 4  
at several temperatures. 
When temperature decreases, the Fermi-surface becomes smaller 
in the staggered-flux phase,
and changes to a point-like Fermi-surface 
discontinuously at the transition temperature 
to the d-wave pairing phase. 
However, this discontinuity is small 
because Fermi surface in the staggered-flux phase has shrunk 
almost to a point at the transition point.
The shape of Fermi-surface in the staggered-flux state of our theory
is segment-like 
because the spectral weight has large enough value 
only along a part of the contour. 

\begin{center}
\begin{figure}[h]
\epsfxsize 7.0cm \epsffile{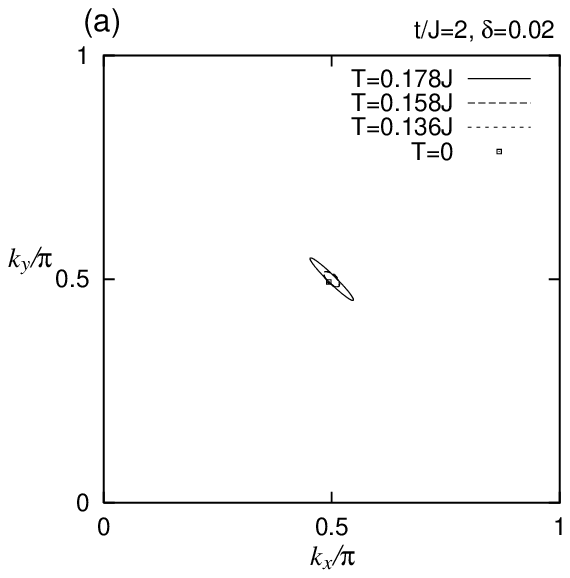}
\epsfxsize 7.0cm \epsffile{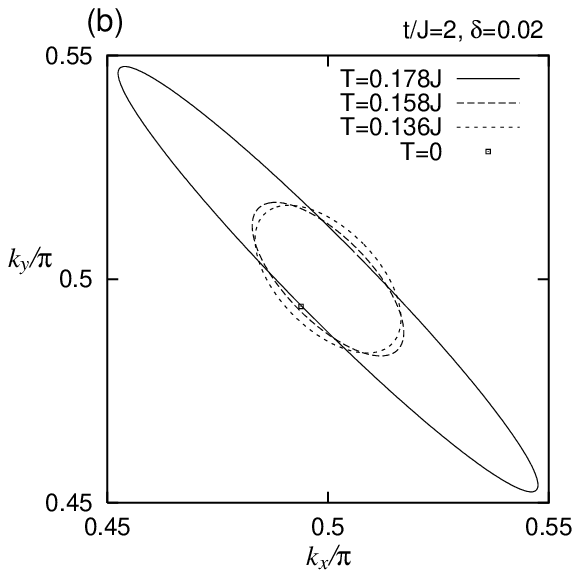}
\caption{
The temperature dependence 
of the zero-energy-line of fermion  at $\delta=0.02$.
In (a), the first quadrant of the Brilluin zone is shown.
In (b), expanded view around ($\pi /2, \pi /2$) is shown. 
In this doping, transition temperature to the d-wave pairing phase 
is about $T=0.135J$. }
\label{FS}
\end{figure}
\end{center}

This result is quite different from the SU(2) mean-field theory, 
where the shape of Fermi-surface is always point-like 
because the particle hole symmetry(i.e.SU(2) symmetry) of fermion remains 
even at finite doping.
As the filling of fermions are always half-filled,
the staggered flux state and d-wave pairing state can be assumed to be gauge equivalent.
Artificial introduction of the phenomenological interactions are needed 
for the fermion 
to have a segment-like Fermi surface in the SU(2) theory. 

Now we give the details. 
The electron Green function $G^{\mathrm e}$ is described by 
the product of spinon Green function $G^{\mathrm s}$ 
and holon Green function $G^{\mathrm h}$ \cite{And88}; 
 \begin{eqnarray}
G^{\mathrm e}(i \omega_n, {\bf k})=
\frac{1}{N \beta} \sum_{ip_\ell, {\bf p}}
G^{\mathrm s}(i p_\ell, {\bf p}) G^{\mathrm h}(i p_\ell- i \omega_n, {\bf p-k}). 
\end{eqnarray}
The contribution of holon Green function comes mainly from
the bottom of the holon band.
The momenta at the bottom of the boson band are 
$(0,0)$ and $(\pi, \pi)$ in the staggered flux state,
and are $(0,0)$, $(\pi, 0)$, 
$(0, \pi)$, and $(\pi, \pi)$ in the $\pi$-flux state.
We can treat the bosons as if they are Bose condensed 
at these band minima. 
In this approximation, 
the spectral function of electron has the following form,
\begin{eqnarray}
A^{\mathrm e}(\omega, {\bf k})&=&\frac{\delta}{2}
\big\{ 
u_{\bf k}^2 \delta( \omega - E^{+}_{\bf k} )
+ |v_{\bf k}|^2  \delta( \omega - E^{-}_{\bf k} ) 
\big\}
\nonumber \\
&&+ A^{in}(\omega, {\bf k}), 
\label{spec}
\end{eqnarray}
where 
$u_{\bf k}^2
= (1+ \epsilon_{\bf k} / ( \epsilon^2_{\bf k}+W^2_{\bf k} )^{1/2} ) /2$,
$|v_{\bf k}|^2
= (1- \epsilon_{\bf k} / ( \epsilon^2_{\bf k}+W^2_{\bf k} )^{1/2} ) /2$,
$A^{in}(\omega, {\bf k})$ 
is the incoherent part of the spectral function, 
$ E^{\pm}_{\bf k}=\pm \sqrt{\epsilon^{2}_{\bf k} + W^{2}_{\bf k} } -\mu $
are spectrums of spinon in the staggered-flux state.
In reality, the boson excitations exist at finite temperature. 
They make the discrete $\delta$-function peaks have a finite width. 
This effect, however, will not change the shape of the Fermi-surface. 
Thus, we show the zero-energy-line of spinon in 
the staggered-flux phase defined by 
$ E^{-}_{\bf k}=- \sqrt{\epsilon^{2}_{\bf k} + W^{2}_{\bf k} } -\mu =0$ 
in Fig.\ref{FS}.  
The upper band spectrum $E^{+}_{\bf k}$ is always positive 
at finite doping where $-\mu \ge 0$.  

\begin{center}
\begin{figure}[h] 
\epsfxsize 7.0cm \epsffile{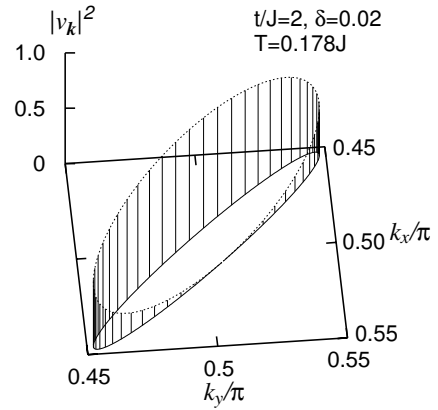}
\caption{
The spectral weight 
$A^{\mathrm e}(\omega, {\bf k})$ 
at $\omega=0$ divided by $\frac{\delta}{2}$
(i.e.  $|v_{\bf k}|^2$) 
at $T=0.178J$. 
View from the ($\pi, \pi$)-direction is shown.  
}
\label{FS2}
\end{figure}
\end{center}

Although the zero-energy-line of fermion forms an ellipse,  
the Fermi-surface can be considered as segment-like. 
The reason is that the intensity of spectrum function 
$\frac{\delta}{2}|v_{\bf k}|^2$ is not symmetric between 
in the inner region(where $|k_{x}|+|k_{y}| \le \pi $) and 
in the outer region(where $|k_{x}|+|k_{y}| \ge \pi $). 
As shown in Fig.\ref{FS2} it is stronger in the inner region. 
For example, 
on the line $k_{x}=k_{y}$, 
where $d_{x^2-y^2}$-gap 
due to staggered-flux order($ W^2_{\bf k}$) is always zero, 
the intensity is $\frac{\delta}{2}$ 
at the inner side of the ellipse as $|v_{\bf k}|^2 =1$,  
but the intensity is zero at the outer side of the ellipse as $|v_{\bf
k}|^2 =0$.

\subsection{\bf Excitation gap at $(0, \pi)$}

\begin{center}
\begin{figure}[h]
\epsfxsize 7.5cm \epsffile{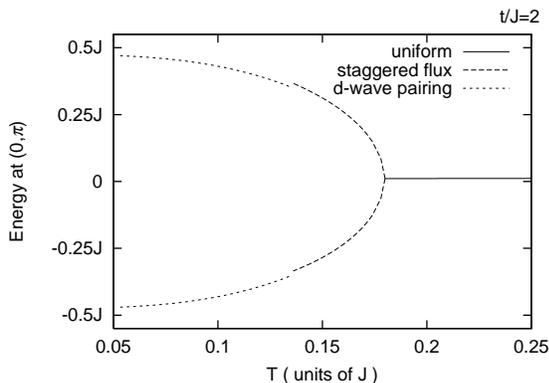}
\caption{
The temperature dependence of the excitation gap at $(0, \pi)$ 
at $\delta=0.02$ for $t/J=2$. 
The phase changes as temperature decreases from the uniform phase 
to the staggered-flux phase and finally to the d-wave pairing phase.
}
\label{gap}
\end{figure}
\end{center}

The temperature dependence of  
the excitation gap at $(0,\pi)$ looks almost continuous at the 
transition point between the staggered-flux phase and the d-wave pairing phase 
(Fig.\ref{gap}),
although it is a first order transition in the present mean field analysis.
The energy at $(0, \pi)$ are 
$E_{\mathrm staggered, \pm}(0, \pi)
= \pm | 4t y_{\mathrm h} + 3Jy_{\mathrm s}/2 | -\mu$  
in the staggered flux state, 
and $E_{\mathrm d-pairing, \pm}(0, \pi) 
= \pm \sqrt{\mu^2 + (3J\eta /2)^2}$
in the d-wave RVB state, respectively.

\subsection{Comparison with experiments}

The ARPES experiments on 
Bi$_{2}$Sr$_{2}$CaCu$_{2}$O$_{8+ \delta}$ (BSCCO)
\cite{Loser96,Marshall96,Ding96,Harris96,Harris97,Ding97,Norman98} 
and recent one on La$_{2-\delta}$Sr$_{\delta}$CuO$_{4}$ (LSCO)\cite{Yoshida0206} 
showed that a segment-like Fermi-surface 
exists near the ($\pi/2, \pi/2$) in pseudo-gap region.
\cite{Loser96,Marshall96,Ding96}. 
There is the gap whose symmetry is $d_{x^2-y^2}$. 
The gap is largest near $(\pi, 0)$ 
and does not exists   
on the $(0, 0)$-$(\pi, \pi)$ line. 
The reason that these Fermi-surface often called segment-like, or arc, 
is due to the shape where the spectral weight is finite. 
It does not form a closed curve.
When one scan on the $(0, 0)$-$(\pi, \pi)$ line, 
the spectral weight only exists in inner region of Brillouin zone
(where $|k_{x}|+|k_{y}| \le \pi$) \cite{Marshall96}.
Thus, this is called segment-like.
When temperature decrease, the segment-like Fermi-surface become 
smaller in pseudo-gap phase and change to the Fermi-point
in the $d_{x^2-y^2}$-wave superconducting state 
\cite{Harris96,Norman98}.  
On the other hand, 
the temperature dependence of the excitation gap at $(0,\pi)$ 
almost looks continuous at the transition point between 
the pseudo-gap phase and 
the $d_{x^2-y^2}$-wave superconducting phase\cite{Norman98}.
Our present results are consistent with these observations. 

%%%%%%%%%%%%%%%%%%%%%%%%%
\section{ Summary }

The competition between the staggered flux, or the d-density wave, and 
the d-wave pairing is analyzed in two-dimensional $t$-$J$ model 
based on U(1) slave boson mean-field theory.
Not only staggered-flux of spinon but also staggered-flux of holon 
are considered, independently. 
In this formalism, 
the hopping order parameter of electron is described by a 
product of hopping order parameters of spinon and holon.
The staggered-flux amplitude of the electron is a difference of 
staggered-flux amplitudes of spinon and holon.
A phase diagram is obtained where
1) a region of the electron staggered-flux state exists,
2) the staggered flux and the d-wave pairing do not coexist, and
3) the ground state is a purely d-wave superconducting state.
In $\pi$-flux phase of spinon, staggered-fluxes of spinon and holon 
cancel completely 
and the staggered-flux order of the electron does not exist.
However, in the staggered-flux phase of spinon whose staggered-flux is not $\pi$, 
fluxes do not cancel completely and staggered-flux order of electron exist. 
Thus, the phase transition between these two phases, 
$\pi$-flux phase and staggered-flux phase of spinon, 
becomes a second order transition in electron picture. 
The order parameter that characterizes this transition 
is the staggered-flux order parameter of the electron. 
The relation between the staggered current of electron and 
that of spinon is provided.
The local cancellation of spinon current and holon current 
does not exclude the possibility of the 
electron staggered current.
It is proved analytically that 
the staggered-flux and the d-wave pairing does not coexist except at half-filling. 
The condition for coexistence of 
staggered-flux and  d-wave pairing as a saddle-point solution is provided.
The instability of these two states, staggered flux and d-wave pairing, 
are discussed. 
The temperature dependence of following two quantities, 
Fermi surface and the excitation gap at $(0,\pi)$, are shown.
The behaviors are consistent with ARPES experiments.
When temperature decreases, 
the segments-like Fermi-surface in the staggered flux phase becomes smaller  
and changes to a point at the transition temperature 
to the d-wave pairing phase.
The temperature dependence of the excitation gap at $(0,\pi)$ 
looks continuous at the transition point between the staggered-flux phase 
and the d-wave pairing phase although these two phases 
are qualitatively different.

%%%%%%%%%%%%%
\section*{Acknowledgements}

K.H. thanks Akihiro Himeda, Takashi Koretsune, Youichi Yanase, 
Masao Ogata, Naoto Nagaosa, Jun-ichiro Kishine, 
 Yousuke Ueno, Kentaro Nomura, Ryuichi Shindou, Shinsei Ryu, 
Yoshifumi Morita, 
and Tetsuo Matsui  
for their useful discussions.  
Numerical computation in this work was partially carried out 
at the Yukawa Institute Computer Facility.

\appendix
%\title{Coexistence condition and instability}
\section*{Coexistence condition and instability}

It is proved analytically that 
the staggered flux and the d-wave pairing 
coexist only at half-filling. 
The instability of these two states, staggered flux and d-wave pairing, 
are also discussed. 

From self-consistent equations (\ref{ys})(\ref{eta})(\ref{yh}), 
following equation is derived. 
It must be satisfied for any states whose free energy  
is at saddle point. 
\bea
0 = y_s \eta \Big[ -(2t)^2 {\sf C} \big( {\sf A + B} \big) +\frac{3J}{4} 
{\sf B} \Big], 
\label{cond}
\eea
where the quantities, {\sf A}, {\sf B}, and {\sf C}, are given by, 
\begin{eqnarray}
{\sf A}&=& \frac{1}{N} \sum_{{\bf k}, s} 
\frac{ \gamma_{{\bf k}-}^{2}}{  2 } 
\frac{ \tanh (\beta E_{{\bf k}s}/2) }{ E_{{\bf k}s}}, 
\\
{\sf B} &=&-\mu \frac{1}{N} \sum_{{\bf k}, s} 
\frac{ \gamma_{{\bf k}-}^{2}}{ 2 ( \epsilon_{ {\bf k} }^2 + W_{ {\bf k} }^2 )^{1/2} } 
\frac{ \tanh (\beta E_{{\bf k}s}/2) }{ E_{{\bf k}s}},  
\\
{\sf C} &=&\frac{1}{N} \sum_{{\bf k}, s} 
\frac{ \gamma_{{\bf k}-}^{2}}{ 2 ( \epsilon_{ {\bf k} }^{{\mathrm h} 2} 
+ W_{ {\bf k} }^{{\mathrm h} 2} )^{1/2} } 
\frac{(-s)}{ {\rm e}^{\beta E^{\mathrm h}_{{\bf k} s}} -1}, 
\end{eqnarray}
where $\gamma_{{\bf k} -} = \cos k_x - \cos k_y$, 
${\sf A}>0$, ${\sf B}\le 0$, ${\sf C} \le 0$.
Here, the quantity {\sf B} is different from $B$ 
which describes the amplitude of the boson-hopping. 
{\sf B} is zero at half-filling as $\mu=0$, 
and is negative at finite doping.
{\sf C} is zero at half-filling or 
when  all holon condense at $(0,0)$ or $(\pi, \pi)$, 
and is negative when excited holon exists 
or when holon condense at $(\pi, 0)$ or  $(0, \pi)$
which is the situation of  $\pi$-flux state.
If the staggered flux and the d-wave pairing 
coexists, $y_{\mathrm{s}}\eta$ is non-zero.  
Then in order for the coexistence, eq.(\ref{cond}) 
requires that the quantity in the square bracket must vanish. 
At half-filling, ${\sf B=C}=0$, so the condition is satisfied.  
At finite doping ${\sf B} <0$, so if ${\sf C}=0$ there is no coexistence.   
When ${\sf C} \ne 0$, we cannot expect the quantity in the square 
bracket vanish except at a special point in the phase space.  
However, from the continuity of the order parameters,
it is impossible to have both $y_{\mathrm{s}}$ and $\eta$ nonzero only at a 
single point.

The curvature of free energy at the saddle points of 
pure staggered-flux state and pure d-wave pairing state 
have the following forms respectively, 
\bea
\frac{\partial^2 F}{\partial y_{\mathrm s}^2}|_{\eta \ne 0, y_{\mathrm s}=0}
&=&  2N \{ -(\frac{3J}{4})^2 {\sf B} + (2t)^2 {\sf C} \},  
\label{cvy}
\\
\frac{\partial^2 F}{\partial \eta^2}|_{y_{\mathrm s}\ne 0, \eta= 0}
&=&2N \big[ \big( \frac{3J}{4} \big)^2 {\sf B} + \frac{3J}{4} 
\{ 1 -  \frac{1}{ 1 - \frac{4(2t)^2}{3J} {\sf C} } \} \big].
\label{cve}
\eea

The stability of the states depends on {\sf B} and {\sf C}. 
There are three cases: 
(i)The case that {\sf B} and {\sf C} are both zero:   
This is realized at half-filling, and  
$ (\partial^2 F/ \partial y_{\mathrm s}^2 )|_{\eta \ne 0, y_{\mathrm s}=0}
= (\partial^2 F/ \partial \eta^2)|_{y_{\mathrm s}\ne 0, \eta= 0}
=0$. 
We need not discuss this case.  
(ii) The case that  {\sf B} is negative and {\sf C} is zero
(this is the same situation that Zhang discussed\cite{Zhang90}): 
In this case, pure d-wave pairing are always stable, 
and pure staggered-flux state is unstable against infinitesimal d-wave pairing.
The curvature at each state are 
$ ( \partial^2 F/ \partial y_{\mathrm s}^2 )|_{\eta \ne 0, y_{\mathrm s}=0}
= - 2N(3J/4)^2 {\sf B} > 0$ and 
$(\partial^2 F/ \partial \eta^2)|_{y_{\mathrm s}\ne 0, \eta= 0}
=2N (3J/4)^2 {\sf B} < 0$.
(iii) The case that {\sf B} and {\sf C} are both negative:
A region where pure d-density wave is stable 
and a region where d-wave pairing is stable 
can both exist.

The situation where {\sf B} is zero and {\sf C} is negative does not occur.
{\sf B} is zero only at half-filling where {\sf C} is zero.

\end{multicols}


\begin{thebibliography}{99}
\bibitem{BC} 
Below the Bose-condensation temperature of holon,
the holon operator can be treated as c-number; 
$b_{i} \rightarrow \langle b_{i} \rangle = z_{i} $, 
where $z_{i}$ is complex c-number. 
Thus, in the U(1) theory the $physical$ electron operator 
can be identified with
the spinon operator;  
$c_{i\sigma}= z^{*}_{i} f_{i\sigma}$, where $|z_{i}|^2=\delta$ 
if all bosons condensate. 
In the SU(2) theory, the electron operator is described by two terms; 
$c_{i\sigma}= z^{*}_{i1}f_{i\sigma}+ 
z^{*}_{i2}\epsilon_{\sigma \sigma^{'}} f^{\dag}_{i \sigma^{'}}$,   
where $z^{*}_{i1}=\langle b^{\dag}_{i1} \rangle$,   
$z^{*}_{i2}=\langle b^{\dag}_{i2} \rangle$,  
and $|z_{i1}|^2+|z_{i2}|^2= \delta$ if all bosons condensate.
The relations between operators, the electron and the spinon,  
are fixed when a direction in the $(z_{i1},z_{i2})$ space is chosen.
%
However, above the Bose-condensation temperature of holon 
where $z_{i}=z_{i1}=z_{i2}=0$, above discussions cannot be used.   
Boson operators cannot be treated as c-number.
Unfortunately, the staggered flux phase exists in this region.
\bibitem{comment_identity}
To be more precise, these identities are not only 
identities between expectation values 
but also identities between operators. 
The reason is following. 
Any correlation functions including these auxiliary fields, 
$\chi_{ij}$, $B_{ij}$, and $\eta_{ij}$, satisfy the identities. 
For example, by differentiating the integrand in (\ref{MFA}) by 
$\chi^{*}_{ij}$, 
any correlation function 
$\langle O_{1} O_{2} ... O_{n} \rangle$ derives to  
$
\langle \chi_{ij} O_{1} O_{2} ... O_{n} \rangle 
- 
\langle f_{i\sigma}^{\dag}f_{j\sigma}  O_{1} O_{2} ... O_{n} \rangle =0.
$
As $O_{1}, O_{2}, ..., O_{n}$ are arbitrary operators, 
this relation means that there is the identy, 
$\chi_{ij}= f_{i\sigma}^{\dag}f_{j\sigma}$, in operator level. 
%It is also operator identities that of the other auxiliary fields, 
The auxiliary fields, 
$B_{ij}$ and $\eta_{ij}$, 
also satisfies the following identities in the operator level,  
$ B_{ij}= b^{\dag}_{i} b_{j}$ and 
$ \eta_{ij} 
= f_{i\uparrow}f_{j\downarrow} 
- f_{i\downarrow}f_{j\uparrow} $, 
respectively.
\bibitem{comment_choice}
Many choices of the $\chi_{ij}$ and the $B_{ij}$ are equivalent. 
These choices of the form of the $\chi_{ij}$ and the $B_{ij}$ corresponds to 
a gauge fixing in the staggered-flux state. 
 The only gauge invariant property of $\chi_{ij}$ and $B_{ij}$ is 
its flux through a plaquette. 
If dimerized phase is excluded, 
%($|\chi_{ij}|=\chi$), 
these auxiliary fields, $\chi_{ij}$ and $B_{ij}$ can parameterized 
by their amplitude and phase, 
$\chi_{ij}=\chi {\mathrm e}^{i \theta_{ij}}$ and 
$B_{ij}=B {\mathrm e}^{i \theta^{\mathrm h}_{ij}}$. 
The flux through a plaquette of spinon(holon) is  
$\theta^{\mathrm (h)}_{ij}+\theta^{\mathrm (h)}_{jk}
+\theta^{\mathrm (h)}_{kl}+\theta^{\mathrm (h)}_{li}$.
In the U(1) slave boson method, electron operator 
$c_{i\sigma}=f_{i \sigma}b^{\dag}_{i}$ is invariant under 
the internal local U(1) transformation, 
$f_{i \sigma} \rightarrow f_{i \sigma} {\mathrm e}^{i \theta_{i}}$ and 
$b_{i } \rightarrow b_{i } {\mathrm e}^{i \theta_{i}}$.
The action is also invariant under this transformation. 
Then auxiliary fields also transform as following,   
$\chi {\mathrm e}^{i \theta_{ij}} \rightarrow 
\chi {\mathrm e}^{ i (\theta_{ij}+ \theta_{i}- \theta_{j}) }$, 
and 
$B {\mathrm e}^{i \theta^{\mathrm h}_{ij}} \rightarrow 
B {\mathrm e}^{ i 
(\theta^{\mathrm h}_{ij}+ \theta_{i}- \theta_{j}) }$.  
However, the flux through a plaquette of $\chi_{ij}$ and that of $B_{ij}$ are 
gauge invariant. 
Of course, the hopping of the electron, 
$ c_{i\sigma}^{\dag}c_{j\sigma}  
=\ {\chi}_{ij}  B_{ij}^{*}$,  
is also gauge invariant quantity, 
$\chi B {\mathrm e}^{ 
i 
(\theta_{ij}- \theta^{\mathrm h}_{ij}) }
\rightarrow 
\chi B {\mathrm e}^{ i 
(\theta_{ij}- \theta^{\mathrm h}_{ij})} 
{\mathrm e}^{ 
i 
\{ (\theta_{i}- \theta_{j})
  -(\theta_{i}- \theta_{j}) \} }
=\chi B {\mathrm e}^{ 
i 
(\theta_{ij}- \theta^{\mathrm h}_{ij}) }
$.
\bibitem{comment_pi} The staggered $\pi$-flux state is equivalent to  
the uniform $\pi$-flux state 
because these states transforms each other by a shift of the $2 \pi$-flux. 
The $2\pi$-flux does not change physics.
\end{thebibliography}
\end{document}